\documentclass[12pt, a4paper]{article}
\usepackage[utf8]{inputenc}
\usepackage{hyperref}
\usepackage{geometry}
\usepackage{graphicx}
\usepackage{subcaption}
\usepackage[numbers]{natbib}
\usepackage{amssymb}                                                                                                                                                                                                                                                                                                                                                                                                                                                                                                                                                                                                                                                                                                                                                                                                                                                                                                                                                                                                                                                                                                                                                                                                                                                                                                                                                                                                                                                                                                                                                                                                                                                                                                                                                                                                                                                                                                                                                                                                                                                                                                                                                                                                                                                                                                                                                                                                                                                                                                                                                                                                                                                                                                                                                                                                                                                                                                                                                                                                                                                                                                                                                                                                                                                                                                                                                                                                                                                                                                                                                                                                                                                                                                                                                                                                                                                                                                                                                                                                                                                                                                                                                                                                                                          
\usepackage{amsmath}
\usepackage{slashed}
\usepackage{anyfontsize}
\usepackage{bm}
\usepackage{slashed}
\usepackage{multirow}
\usepackage{float}
\usepackage[font=small]{caption}
\usepackage{listings}
\newgeometry{lmargin=1.1in, rmargin=1.1in, tmargin=1in, bmargin=1in}
\title{The study of conformal geometry and its exact solution of the geodesic deviation equation}
\author{B.T.T.Wong\footnote{CERN, u3500478@connect.hku.hk}}
\date{}

\begin{document}

\maketitle
\begin{abstract}
In this paper, the geometric properties of the conformal metric are studied and its exact solution of the geodesic deviation equation is presented. We also find out the stress-energy tensor of this geometry and compare it with the usual prefect-fluid case, obtaining an equation of state as $P = -\frac{1}{3}\rho$ in 4D space-time dimension. Finally, the low-energy regime of the metric is studied, in which we obtain the stress-energy tensor proportional to the projection tensor.
\end{abstract}

\section{Introduction}
The Einstein field equation is a non-linear, tensorial, second ordered partial differential equation which is very difficult to solve. The seek for exact solution under special cases with symmetry condition have been preformed throughout the decades. For example, the famous Schwarzschild solution for a black hole \cite{Sch}, and numerous fruitful solutions to different cases of vacuum Einstein field equation, for which the stress-energy tensor $T_{\mu\nu} =0$. This includes, for instance, Reissner–Nordström electrovacuum, Kerr–Newman electrovacuum, Melvin electrovacuum, Weyl–Maxwell electrovacuum, Bertotti–Robinson electrovacuum and so on \cite{1,book}. 

Solving and analyzing the geodesic equation gives the dynamics of equation of motion of the system. For example, the famous case of the precession of the perihelion of Mercury \cite{Mercury, Weinberg}. Solving the geodesic deviation  equation (acceleration equation) enables one to understand the equation of motion of nearby test-particles under gravitational effect. For example, solving the linearlized, weak-field limit equation gives the gravitational wave solution \cite{Einstein,Gravitation}. 

In this paper, we are interested in studying the geometry properties of the conformal metric and solve the exact solution of its corresponding geodesic deviation equation. Conformal transformation invariance is of extreme importance in string theory, where the Polyakov action is conformal transformatin of the metric\cite{4},
\begin{equation}
g_{\mu\nu}^\prime (x^\prime) = e^{2\Omega(x)} g_{\mu\nu} (x)\,.
\end{equation}
A particular gauge choice is to choose the original metric as the flat metric such that $g_{\mu\nu}(x) = \eta_{\mu\nu}$, known as the Weyl gauge. This choice is of particular importance and it is used in the Polyakov bosonic action. Due to its significance in different areas in physics, it is essential to study the conformal geometry in details.

\section{Geometric properties of conformal metric} 
In conformal geometry, here we consider the metric in the generic form as follow
\begin{equation}
g_{\mu\nu} (x) = e^{a\theta(x)}\eta_{\mu\nu} \,,
\end{equation}
where $a=\pm 1$ is called the deterministic parameter. If $a= +1$, we have the exponential growth of the flat background metric; if $a=-1$, we have the exponential decay of the flat background metric. Next we would like to find out all the geometric quantities, and we will work in the general $D$ dimensions instead of just $D=4$. This is to facilitate the study of dimensional analysis. 

The Christoffel connection is given by
\begin{equation}
\Gamma^{\rho}_{\mu\nu} = \frac{1}{2} g^{\rho\sigma} ( \partial_{\mu} g_{\sigma\nu} + \partial_{\nu}g_{\sigma \mu} - \partial_{\sigma}g_{\mu\nu} ) \,.
\end{equation}
The Christoffel connection of this conformal metric is
\begin{equation} \label{eq:ChristoffelConformal}
\Gamma^{\rho}_{\mu\nu} = \frac{a}{2} ( \delta^{\rho}_{\nu}\partial_{\mu}\theta + \delta^{\rho}_{\mu}\partial_{\nu}\theta  - \eta_{\mu\nu}\partial^{\rho}\theta ) \,. 
\end{equation}
Therefore, the geodesic equation is
\begin{equation}
\frac{d^2 x^{\rho}}{d\tau^2} + \Gamma^{\rho}_{\mu\nu} \frac{dx^{\mu}}{d\tau }\frac{dx^{\nu}}{d\tau } = \frac{d^2 x^{\rho}}{d\tau^2}+  \frac{a}{2} ( \delta^{\rho}_{\nu}\partial_{\mu}\theta + \delta^{\rho}_{\mu}\partial_{\nu}\theta  - \eta_{\mu\nu}\partial^{\rho}\theta ) \frac{dx^{\mu}}{d\tau }\frac{dx^{\nu}}{d\tau } = 0 \,.
\end{equation}
Next, the Riemanian curvature tensor is given by
\begin{equation}
R^{\rho}_{\,\,\,\sigma\mu\nu} = \partial_{\mu} \Gamma^{\rho}_{\nu\sigma} - \partial_{\nu} \Gamma^{\rho}_{\mu\sigma} + \Gamma^{\rho}_{\mu\lambda} \Gamma^{\lambda}_{\sigma \nu} - \Gamma^{\rho}_{\nu\lambda} \Gamma^{\lambda}_{\sigma \mu} \,.
\end{equation}
Therefore one can compute first two terms for the curvature tensor as
\begin{equation}
\partial_{\mu}\Gamma^{\rho}_{\nu\sigma}- \partial_{\nu}\Gamma^{\rho}_{\mu\sigma} = \frac{a}{2} ( \delta^{\rho}_{\nu}\partial_{\mu}\partial_{\sigma}\theta - \delta^{\rho}_{\mu} \partial_{\nu}\partial_{\sigma}\theta - \eta_{\nu\sigma}\partial_{\mu}\partial^{\rho}\theta - \eta_{\mu\sigma}\partial_{\nu}\partial^{\rho} \theta ) \,.
\end{equation}
Then we compute the third term,
\begin{equation}
\begin{aligned}
\Gamma^{\rho}_{\mu\lambda}\Gamma^{\lambda}_{\nu\sigma} & =  \frac{a^2}{4} \big( \delta^{\rho}_{\lambda}\delta^{\lambda}_{\sigma}\partial_{\mu}\theta \partial_{\nu}\theta + \delta^{\rho}_{\lambda} \delta^{\lambda}_{\nu}\partial_{\mu}\theta \partial_{\sigma}\theta - \delta^{\rho}_{\lambda}\eta_{\nu\sigma} \partial_{\mu}\theta \partial^{\lambda}\theta \\
& \,\,\,\, + \delta^{\rho}_{\mu}\delta^{\lambda}_{\sigma}\partial_{\lambda}\theta \partial_{\nu}\theta + \delta^{\rho}_{\mu} \delta^{\lambda}_{\nu}\partial_{\lambda}\theta  \partial_{\sigma} \theta - \delta^{\rho}_{\mu} \eta_{\nu\sigma}\partial_{\lambda}\theta \partial^{\lambda}\theta \\
& \,\,\,\, + \delta^{\lambda}_{\sigma} \eta_{\mu\lambda} \partial_{\mu}\theta \partial^{\rho}\theta - \delta^{\lambda}_{\nu}\eta_{\mu\lambda}\partial^{\rho}\theta \partial_{\sigma}\theta + \eta_{\mu\lambda} \eta_{\nu\sigma} \partial^{\rho}\theta \partial^{\lambda}\theta \big) \\
&= \frac{a^2}{4} \big( \delta^{\rho}_{\sigma} \partial_{\mu}\theta \partial_{\nu}\theta + \delta^{\rho}_{\nu} \partial_{\mu}\theta \partial_{\sigma}\theta - \eta_{\nu\sigma}\partial_{\mu}\theta \partial^{\rho}\theta \\
&\,\,\,\, + \delta^{\rho}_{\mu} \partial_{\sigma}\theta \partial_{\nu}\theta + \delta^{\rho}_{\mu}\partial_{\nu}\theta \partial_{\sigma}\theta - \delta^{\rho}_{\mu} \eta_{\nu\sigma} \partial_{\lambda}\theta \partial^{\lambda}\theta \\
& \,\,\,\,- \eta_{\mu\sigma} \partial_{\nu}\theta \partial^{\rho}\theta - \eta_{\mu\nu}\partial^{\rho}\theta \partial_{\sigma}\theta + \eta_{\nu\sigma}\partial^{\rho}\theta \partial_{\mu}\theta \big) \,, \\
\end{aligned}
\end{equation}
The third term and the last term cancel, and the forth term and the fifth term combine, thus we get
\begin{equation} \label{eq:step1}
\Gamma^{\rho}_{\mu\lambda}\Gamma^{\lambda}_{\nu\sigma} = \frac{a^2}{4} \big( \delta^{\rho}_{\sigma} \partial_{\mu}\theta \partial_{\nu}\theta + \delta^{\rho}_{\nu}\partial_{\mu}\theta \partial_{\sigma}\theta + 2\delta^{\rho}_{\mu}\partial_{\nu}\theta \partial_{\sigma}\theta  - \delta^{\rho}_{\nu} \eta_{\mu\sigma} \partial_{\lambda}\theta \partial^{\lambda}\theta - \eta_{\mu\sigma} \partial_{\nu}\theta \partial^{\rho}\theta - \eta_{\mu\nu}\partial^{\rho}\theta \partial_{\sigma}\theta  \big) \,.
\end{equation} 
Now by swapping $\mu$ and $\nu$ index we obtain the other term as
\begin{equation} \label{eq:step2}
\Gamma^{\rho}_{\nu\lambda}\Gamma^{\lambda}_{\mu\sigma} = \frac{a^2}{4} \big( \delta^{\rho}_{\sigma} \partial_{\nu}\theta \partial_{\mu}\theta + \delta^{\rho}_{\mu}\partial_{\nu}\theta \partial_{\sigma}\theta + 2\delta^{\rho}_{\nu}\partial_{\mu}\theta \partial_{\sigma}\theta  - \delta^{\rho}_{\mu} \eta_{\nu\sigma} \partial_{\lambda}\theta \partial^{\lambda}\theta - \eta_{\nu\sigma} \partial_{\mu}\theta \partial^{\rho}\theta - \eta_{\nu\mu}\partial^{\rho}\theta \partial_{\sigma}\theta  \big) \,.
\end{equation}
After some algebra cancellation and combining all the results, finally we obtain the Riemannian curvature tensor as
\begin{equation} \label{eq:CurvatureTensorResult}
\begin{aligned}
R^{\rho}_{\,\,\,\sigma \mu\nu} &= \frac{a}{2} \big[ ( \delta^{\rho}_{\nu} \partial_{\mu} - \delta^{\rho}_{\mu} \partial_{\nu}) \partial_{\sigma}\theta - ( \eta_{\nu\sigma}\partial_{\mu} - \eta_{\mu\sigma}\partial_{\nu} ) \partial^{\rho}\theta  \big] \\
& \,\,\,\, +\frac{a^2}{4} \big[ (\delta^{\rho}_{\mu} \partial_{\nu}\theta  - \delta^{\rho}_{\nu}\partial_{\mu}\theta )\partial_{\sigma}\theta - (\eta_{\mu\sigma} \partial_{\nu}\theta - \eta_{\nu\sigma}\partial_{\mu}\theta  ) \partial^{\rho}\theta \\
& \,\,\,\, - (\delta^{\rho}_{\mu}\eta_{\nu\sigma} - \delta^{\rho}_{\nu}\eta_{\mu\sigma} ) \partial_{\lambda}\theta \partial^{\lambda} \theta \big] \,. \\
\end{aligned}
\end{equation}
Then using the fact that $g^{\sigma\nu}\eta_{\sigma\nu} = e^{-a\theta} \eta^{\sigma\nu} \eta_{\sigma\nu} = De^{-a \theta}$, and the fact that $\delta^{\rho}_{\rho} = D$, the Ricci scalar is computed as the contraction of the first and third index, which gives

\begin{equation}
\begin{aligned}
R_{\sigma\nu} &= \frac{a}{2} \big[ ( \delta^{\rho}_{\nu} \partial_{\rho} - \delta^{\rho}_{\rho} \partial_{\nu}) \partial_{\sigma}\theta - ( \eta_{\nu\sigma}\partial_{\rho} - \eta_{\rho\sigma}\partial_{\nu} ) \partial^{\rho}\theta  \big] \\
& \,\,\,\, +\frac{a^2}{4} \big[ (\delta^{\rho}_{\rho} \partial_{\nu}\theta  - \delta^{\rho}_{\nu}\partial_{\rho}\theta )\partial_{\sigma}\theta - (\eta_{\rho\sigma}\partial_{\nu}\theta - \eta_{\nu\sigma}\partial_{\rho}\theta  ) \partial^{\rho}\theta \\
& \,\,\,\, - (\delta^{\rho}_{\rho}\eta_{\nu\sigma} - \delta^{\rho}_{\nu}\eta_{\rho\sigma} ) \partial_{\lambda}\theta \partial^{\lambda} \theta \big]  \\
&= \frac{a}{2} \big[ (\partial_{\nu} - D\partial_{\nu}) \partial_{\sigma}\theta -(\eta_{\nu\sigma} \Box \theta - \partial_{\sigma} \partial_{\nu}\theta )  \big] \\
& \,\,\,\, + \frac{a^2}{4} \big[ D\partial_{\nu}\theta - \partial_{\nu}\theta)\partial_{\sigma}\theta- (\partial_{\sigma}\theta \partial_{\nu}\theta - \eta_{\nu\sigma} ) \partial_{\rho}\theta \partial^{\rho} \theta \\
& \,\,\,\, - (D\eta_{\nu\sigma}- \eta_{\nu\sigma})\partial_{\rho}\theta \partial^{\rho}\theta  \big] \\
& = \frac{a}{2} \big[ (1-D) \partial_{\nu} \partial_{\sigma}\theta - \eta_{\nu\sigma}\Box\theta + \partial_{\sigma}\partial_{\nu}\theta  \big] \\
& \,\,\,\, +\frac{a^2}{4} \big[ (D-1) \partial_{\nu}\theta \partial_{\sigma}\theta -\partial_{\sigma}\theta \partial_{\nu}\theta + \eta_{\nu\sigma}\partial_{\rho}\theta \partial_{\rho}\theta -D \eta_{\nu\sigma} \partial_{\rho}\theta \partial^{\rho}\theta + \eta_{\nu\rho}\partial_{\rho}\theta \partial^{\rho}\theta  \big] \\
& = \frac{a}{2} \big[ (2-D) \partial_{\nu}\partial_{\sigma}\theta - \eta_{\nu\sigma}\Box\theta  \big] + \frac{a^2}{4} \big[ (D-2)\partial_{\nu}\theta\partial_{\sigma}\theta - (D-2)\eta_{\nu\sigma} \partial_{\rho}\theta \partial^{\rho}\theta    \big] \\
\end{aligned}
\end{equation}
Therefore finally we have the Ricci tensor as
\begin{equation} \label{eq:Riccitensor}
R_{\sigma\nu} = -\frac{a}{2}\big[ \eta_{\nu\sigma} \Box \theta + (D-2) \partial_{\nu} \partial_{\sigma}\theta  \big] - \frac{D-2}{4}a^2 ( \eta_{\nu\sigma} \partial_{\rho}\theta \partial^{\rho}\theta - \partial_{\nu}\theta\partial_{\sigma}\theta   )  \,.
\end{equation}
Then using the fact that $g^{\sigma\nu}\eta_{\sigma\nu} = e^{-ia\theta} \eta^{\sigma\nu} \eta_{\sigma\nu} = De^{-ia \theta}$, the Ricci scalar is computed as
 \begin{equation} \label{eq:RicciScalar}
R= -a(D-1)e^{-a\theta} \Box \theta - \frac{1}{4}(D-1)(D-2) a^2 e^{-a\theta}\partial_{\rho}\theta\partial^{\rho}\theta \,.
\end{equation}

After studying the general geometrical properties, next we are in particular interested in $\theta(x)$ as a linear function of $x^{\mu}$,
\begin{equation}
\theta = k_{\mu}x^{\mu} + c \,,
\end{equation}
where $c$ is some constant, and we can always choose it to be zero for convenience. This gives
\begin{equation} \label{eq:Spacetimeoscillation}
g_{\mu\nu} = e^{ak\cdot x} \eta_{\mu\nu} \,,
\end{equation}
which describes the exponential growth or decay of the flat background metric. 

Using equation (\ref{eq:ChristoffelConformal}), we find that the Christoffel connection is
\begin{equation}
\Gamma^\rho_{\mu\nu} = \frac{a^2}{2} ( k_\mu  \delta^\rho_\nu + k_\nu \delta^\rho_\mu -k^\rho \eta_{\mu\nu} ) \,.
\end{equation} 
Explicitly, component-wise we have
\begin{equation*} \label{eq:christoffelk}
\Gamma^{0}_{00} =\Gamma^{0}_{ii}= \frac{a^2 \omega}{2} \quad , \quad \Gamma^{i}_{i0} = \Gamma^{i}_{0i} = -\frac{a^2}{2} k^i \quad ,  \quad \Gamma^{0}_{ij} =0 \quad ,
\end{equation*} 
\begin{equation*}
\Gamma^{i}_{00} = -\frac{a^2 k^i}{2} \quad , \quad  \Gamma^{i}_{ii} =-\frac{a^2 k^i}{2}  \quad , \quad  \Gamma^{i}_{i0} = \Gamma^{i}_{0i} = \frac{a^2 \omega}{2} \quad , \quad  \Gamma^{i}_{ij} = \Gamma^{i}_{ji} = -\frac{a^2 k^j}{2} \quad ,
\end{equation*}
\begin{equation}
\Gamma^{i}_{jj} = \frac{a^2 k^i}{2} \quad , \quad  \Gamma^{i}_{jk} = 0 \,\,\,\, \text{for} \,\, i\neq j\neq k \,.
\end{equation}
It is noted that we have used the fact of  $k_i  = -k^i$. Now we can obtain the temporary-component geodesic equation as follow,
\begin{equation}
\frac{d^2 x^0}{d\tau^2} + \Gamma^{0}_{00} \frac{dx^0}{d\tau}\frac{dx^0}{d\tau} + \Gamma^0_{ii}\frac{dx^i}{d\tau}\frac{dx^i}{d\tau} + 2\Gamma^{0}_{0i} \frac{dx^0}{d\tau} \frac{dx^i}{d\tau} = 0 \,,
\end{equation}
explicitly,
\begin{equation} \label{eq:geodesic1}
\frac{d^2 t}{d\tau^2} + \frac{a^2 \omega}{2} \bigg[ \bigg(\frac{dt}{d\tau}\bigg)^2 +  \bigg(\frac{dx}{d\tau}\bigg)^2 +  \bigg(\frac{dy}{d\tau}\bigg)^2 +  \bigg(\frac{dz}{d\tau}\bigg)^2  \bigg] + a^2 \frac{dt}{d\tau}\bigg( k_x \frac{dx}{d\tau} + k_y \frac{dy}{d\tau} + k_z \frac{dz}{d\tau}  \bigg) = 0\,.
\end{equation}
For the spatial component,
\begin{equation}
\frac{d^2 x^i}{d\tau^2} + \Gamma^{i}_{00} \frac{dx^0}{d\tau}\frac{dx^0}{d\tau} + \Gamma^i_{ii}\frac{dx^i}{d\tau}\frac{dx^i}{d\tau} + 2\Gamma^{i}_{0i} \frac{dx^0}{d\tau} \frac{dx^i}{d\tau} + 2\Gamma^{i}_{ji} \frac{dx^j}{d\tau} \frac{dx^i}{d\tau} = 0 \,,
\end{equation}
where for the last term $j \neq i$. Explicitly for the x-component,
\begin{equation} \label{eq:geodesic2}
\begin{aligned}
&\frac{d^2 x}{d\tau^2}  -\frac{a^2 k_x}{2} \bigg[  \bigg(\frac{dt}{d\tau} \bigg)^2 + \bigg(\frac{dx}{d\tau} \bigg)^2 - \bigg(\frac{dy}{d\tau} \bigg)^2 - \bigg(\frac{dz}{d\tau} \bigg)^2\bigg]  \\
&\quad\quad\quad\quad\quad\quad\quad\quad\quad\quad\quad\quad\quad + a^2 \frac{dx}{d\tau} \bigg( \omega \frac{dt}{d\tau} - k_y \frac{dy}{d\tau} - k_z  \frac{dz}{d\tau}  \bigg) = 0 \,. 
\end{aligned}
\end{equation}
For the $y$-component, 
\begin{equation} \label{eq:geodesic3}
\begin{aligned}
&\frac{d^2 y}{d\tau^2}  -\frac{a^2 k_y}{2} \bigg[  \bigg(\frac{dt}{d\tau} \bigg)^2 + \bigg(\frac{dy}{d\tau} \bigg)^2 - \bigg(\frac{dx}{d\tau} \bigg)^2 - \bigg(\frac{dz}{d\tau} \bigg)^2\bigg]   \\
&\quad\quad\quad\quad\quad\quad\quad\quad\quad\quad\quad\quad\quad + a^2 \frac{dy}{d\tau} \bigg( \omega \frac{dt}{d\tau}  -  k_x \frac{dx}{d\tau} - k_z  \frac{dz}{d\tau} \bigg) = 0 \,. 
\end{aligned}
\end{equation}
For the $z$-component,
\begin{equation} \label{eq:geodesic4}
\begin{aligned}
&\frac{d^2 z}{d\tau^2}  -\frac{a^2 k_z}{2} \bigg[  \bigg(\frac{dt}{d\tau} \bigg)^2 + \bigg(\frac{dz}{d\tau} \bigg)^2 - \bigg(\frac{dx}{d\tau} \bigg)^2 - \bigg(\frac{dy}{d\tau} \bigg)^2\bigg]   \\
&\quad\quad\quad\quad\quad\quad\quad\quad\quad\quad\quad\quad\quad + a^2 \frac{dz}{d\tau} \bigg( \omega \frac{dt}{d\tau}  -  k_x \frac{dx}{d\tau} - k_y  \frac{dy}{d\tau} \bigg) = 0 \,. 
\end{aligned}
\end{equation}
The four equations (\ref{eq:geodesic1}), (\ref{eq:geodesic2}), (\ref{eq:geodesic3}) and (\ref{eq:geodesic4}) are a set of coupled, second order non-linear differential equations which is extremely difficult to solve.
Also note that as $a^2 =1$, therefore no matter the case for exponential growth or decay of the flat metric, we will get the same set of geodesic equations.

The linear form of the phase has important consequences for the the geometry. Using the general result obtained in (\ref{eq:CurvatureTensorResult}), the Riemannian curvature tensor becomes
 \begin{equation}\label{eq:CurvatureTensor}
R^{\rho}_{\,\,\,\sigma \mu\nu} = \frac{a^2}{4} \big[ \delta^{\rho}_{\mu} k_{\nu}k_{\sigma} - \delta^{\rho}_{\nu} k_{\mu} k_{\sigma} - \eta_{\mu\sigma} k_{\nu} k^{\rho} + \eta_{\nu\sigma}k_{\mu} k^{\rho} - ( \delta_{\mu}^\rho \eta_{\nu\sigma} - \delta^{\rho}_{\nu} \eta_{\mu\sigma} ) k^2 \big] \,.
\end{equation}
The Ricci tensor becomes,
\begin{equation} \label{eq:RicciProjectionTensor}
R_{\mu\nu} = -\frac{D-2}{4}a^2 ( k^2 \eta_{\mu\nu} - k_{\mu}k_{\nu} ) \,,
\end{equation}
which is proportional to the projection tensor. Therefore, The Ricci scalar is
\begin{equation} \label{eq:RicciScalar}
R= -\frac{(D-1)(D-2)}{4} a^2 k^2 e^{-ak \cdot x} \,.
\end{equation} 
Then the Einstein tensor is given by
\begin{equation} \label{eq:EinsteinTensor}
\begin{aligned}
G_{\mu\nu} &= R_{\mu\nu} - \frac{1}{2}Rg_{\mu\nu } \\ 
&= -\frac{D-2}{4}a^2 \bigg( (k^2 \eta_{\mu\nu}-k_{\mu}k_{\nu}) -\frac{D-1}{2} k^2 \eta_{\mu\nu}  \bigg)  \,.
\end{aligned}
\end{equation}
It can be checked that by direct computation the Riemannian tensor in 2D vanishes. We can also see that when $D=2$, the Ricci tensor and Ricci scalar vanish. It follows that the Einstein tensor $G_{\mu\nu}$ vanishes. Thus for $D=2$, the conformal spacetime geometry has all vanishing geometry quantities and hence considered as flat.

\section{Stress-energy tensor of linearlized conformal metric}
In this section we would like to study the source of Einstein tensor for our conformal geometry in 4D spacetime. According to equations  (\ref{eq:RicciProjectionTensor}), (\ref{eq:RicciScalar}) and the Einstein field equation (\ref{eq:EinsteinTensor}), we have
\begin{equation}
G_{\mu\nu}= -\frac{1}{2}a^2 (k^2 \eta_{\mu\nu} - k_{\mu}k_{\nu}) + \frac{3}{4}a^2 e^{-ak\cdot x} k^2 g_{\mu\nu} = 8\pi G T_{\mu\nu}\,.
\end{equation}
Substituting $g_{\mu\nu} = e^{ak\cdot x} \eta_{\mu\nu}$ back then we obtain the stress-energy tensor as follow,
\begin{equation} \label{eq:StressEnergyTensorU(1)spacetimeManifold}
T_{\mu\nu} = \frac{a^2}{16\pi G} \big( k_{\mu}k_{\nu} + \frac{1}{2} k^2   \eta_{\mu\nu}  \big)\,,
\end{equation}
It follows that the stress energy scalar given by $T=g^{\mu\nu} T_{\mu\nu}$ or by using the identity $R=-8\pi GT$ we have
\begin{equation}
T = \frac{3m^2 a^2}{16 \pi G} e^{-ak \cdot x}\,.
\end{equation}
Hence the stress-energy scalar is an exponentially decreasing or growing function proportional to $k^2=m^2$. This form of energy momentum tensor is comparable to the standard one of perfect fluid, which is given by \cite{Hawking}
\begin{equation}
T_{\mu\nu} = ( \rho + P ) u_{\mu} u_{\nu} - P\eta_{\mu\nu} \,.
\end{equation}
Using the relation of $p^{\mu} = mu^{\mu} = \hbar k^{\mu}$ (we will use natural units here by taking $\hbar =1$), then our energy momentum tensor reads
\begin{equation}
T_{\mu\nu} = \frac{a^2}{16\pi G}( m^2 u_{\mu} u_{\nu} + \frac{1}{2} m^2 \eta_{\mu\nu} ) \,.
\end{equation}
Compare terms with the fluid energy-momentum tensor, we identify the pressure as
\begin{equation} \label{eq:Pressure}
P = -\frac{m^2 a^2}{32 \pi G} \,,
\end{equation}
and
\begin{equation}
\rho + P = \frac{m^2 a^2}{16 \pi G} \,.
\end{equation}
Therefore we obtain the density as
\begin{equation} \label{eq:density}
\rho = \frac{3m^2 a^2}{32 \pi G} \,,
\end{equation}
By equations (\ref{eq:Pressure}) and (\ref{eq:density}), we obtain the equation of state (E.O.S.) as
\begin{equation}
P =-\frac{1}{3} \rho \,,
\end{equation}
which corresponds to $w= -\frac{1}{3}$ for the general equation of state $P = w \rho c^2$. The matter density can be expressed in terms of the stress-energy scalar as
\begin{equation}
T = \rho e^{- a k \cdot x} \,.
\end{equation} 
Next we would like to study for the case of virtual particle $k^2 < 0$ .  Then we have
\begin{equation}
P = \frac{|k^2| a^2}{32 \pi G} > 0 \,\,\,\,{\text{and}}\,\,\,\, \rho = \frac{-3|k^2| a^2}{32 \pi G} < 0 \,.
\end{equation}
For the free photon case $k^2 = m^2 =0$, then we will have
\begin{equation}
k_{\mu} k_{\nu} = -16\pi G T_{\mu\nu}\,.
\end{equation}
Contracting both sides by the inverse flat metric,  the L.H.S vanishes, thus 
\begin{equation}
T=0 \,,
\end{equation}
therefore a free photon corresponds to vanishing stress-energy scalar.

\section{Geometric properties of conformal metric in low energy-momentum limit}
In this section we study the properties for the conformal metric  in low energy-momentum regime in which the perturbation theory can apply. In perturbation theory, the metric $g_{\mu\nu}$ can be separated into a flat background metric and a perturbation metric $h_{\mu\nu}$,
\begin{equation}
g_{\mu\nu} = \eta_{\mu\nu} + h_{\mu\nu} \,.
\end{equation}
Perturbatively, the conformal metric can be expanded as
\begin{equation}
g_{\mu\nu} = \eta_{\mu\nu} + a (k\cdot x) \eta_{\mu\nu} + \cdots \,,
\end{equation}
which amounts to the study of low energy-momentum regime of the theory that is considered as weak perturbation of the flat metric background.
We would study the first and second order term of the perturbation series of the metric. The idea is to consider the metric that can be decomposed into a flat metric with a small perturbation $h_{\mu\nu}$, so we will use the linearized perturbation theory of general relativity \cite{Gravitation}.
We just want to find all the geometry quantities that are first order in $h_{\mu\nu}$.  Up to the first order in $h_{\mu\nu}$ next we compute the Christoffel connection
\begin{equation}
\Gamma^{\rho}_{\mu\nu} = \frac{1}{2} \eta^{\rho\lambda}(\partial_{\mu}h_{\nu\lambda} + \partial_{\nu}h_{\lambda\mu} - \partial_{\lambda}h_{\mu\nu} ) 
\end{equation}
and the Riemannian curvature tensor is
\begin{equation}
R^{\rho}_{\,\,\,\sigma\mu\nu} = \partial_{\mu} \Gamma^{\rho}_{\nu\sigma} - \partial_{\nu}\Gamma^{\rho}_{\mu\sigma} \,.
\end{equation}
There are no terms involving two products of Christoffel connection because we only consider the first order in $h_{\mu\nu}$. The Ricci tensor in linearized GR is
\begin{equation}
R_{\mu\nu} = \frac{1}{2}( \partial_{\sigma} \partial_{\nu} h^{\sigma}_{\,\,\,\mu} + \partial_\sigma \partial_\mu h^{\sigma}_{\,\,\,\nu} - \partial_{\mu}\partial_{\nu}h -\Box h_{\mu\nu} ) \,,
\end{equation}
and the Ricci scalar is
\begin{equation}
R = \partial_{\mu}\partial_{\nu}h^{\mu\nu} - \Box h \,.
\end{equation}
The Einstein tensor in linearized perturbation GR is \cite{Gravitation}
\begin{equation}
G_{\mu\nu}= R_{\mu\nu} - \frac{1}{2}\eta_{\mu\nu}R=  \frac{1}{2}( \partial_{\sigma} \partial_{\nu} h^{\sigma}_{\,\,\,\mu} + \partial_\sigma \partial_\mu h^{\sigma}_{\,\,\,\nu} - \partial_{\mu}\partial_{\nu}h -\Box h_{\mu\nu} -\eta_{\mu\nu}\partial_{\rho}\partial_{\lambda}h^{\rho\lambda} + \eta_{\mu\nu}\Box h )\,.
\end{equation}
Now we would apply the above perturbation theory for our case. Recall
\begin{equation}
 g_{\mu\nu }(x)  = \eta_{\mu\nu} + a (k\cdot x) \eta_{\mu\nu}+ \frac{a^2 (k \cdot x)^2}{2}\eta_{\mu\nu} + \cdots = \eta_{\mu\nu} + h_{\mu\nu}+ u_{\mu\nu} + \cdots 
\end{equation}
with small $k_\rho <<1$. Then we have the first order term as
\begin{equation}
h_{\mu\nu} = a k_{\rho} x^\rho \eta_{\mu\nu} \,.
\end{equation}
Thus we consider small $k$, i.e. locally small oscillation mode. Since terms like $\partial_{\mu} h_{\nu\lambda} = a k_\mu \eta_{\nu\lambda} $ would vanish when subjected to second order partial differentiation, therefore the first order term in the series does not give any contribution to the curvature. Thus we consider the next order term $u_{\mu\nu}= \frac{a^2}{2} k_{\rho}k_\sigma x^\rho x^\sigma \eta_{\mu\nu} $\,. Then we have $\partial_{\alpha} h_{\mu\nu} = a^2 k_{\alpha}k_{\sigma}x^\sigma \eta_{\mu\nu}$ which is not a constant. Thus the non-vanishing Christoffel connection is
\begin{equation}
\Gamma^\rho_{\mu\nu} = \frac{a^2}{2} (k \cdot x ) ( k_\mu  \delta^\rho_\nu + k_\nu \delta^\rho_\mu -k^\rho \eta_{\mu\nu} ) \,,
\end{equation}
which is similar to the original one in equation \ref{eq:christoffelk}, expect for the extra phase factor $(k\cdot x)$. The geodesic equations will be similar to that of the original case in section 2, except with the front phase factor $(k\cdot x)$. Then the Riemannian curvature tensor is
\begin{equation}
\begin{aligned}
R^{\rho}_{\,\,\,\sigma\mu\nu} &= \partial_{\mu} \Gamma^{\rho}_{\nu\sigma} - \partial_{\nu}\Gamma^{\rho}_{\mu\sigma} \\
&= \frac{a^2}{2} \big[ \partial_{\mu} (k\cdot x) (k_\nu \delta^\rho_\sigma +k_\sigma \delta^\rho_\nu - k^\rho\eta_{\nu\sigma} )-  \partial_{\nu} (k\cdot x) (k_\mu \delta^\rho_\sigma +k_\sigma \delta^\rho_\mu - k^\rho\eta_{\mu\sigma} )  \big] \\
&=\frac{a^2}{2} \big[ k_{\mu}k_{\nu} \delta^{\rho}_{\sigma} + k_{\mu}k_{\sigma} \delta^{\rho_\nu} - k_{\mu}k^\rho \eta_{\nu\sigma} - k_{\nu}k_{\mu} \delta^{\rho}_\sigma -k_\nu k_\sigma \delta^\rho_\mu + k_\nu k^\rho \eta_{\mu\sigma}\big] \\
&= \frac{a^2}{2} \big[ k_\sigma (k_\mu \delta^\rho_\nu - k_\nu \delta^\rho_\mu ) + k^\rho(k_\nu \eta_{\mu\sigma} - k_\mu \eta_{\nu\sigma}) \big] \,.
\end{aligned}
\end{equation}
Next we calculate the Ricci tensor for low $k$ regime,
\begin{equation}
R_{\sigma\nu} = -\frac{a^2}{2} \Big( k^2\eta_{\nu\sigma} +(D-2)k_{\sigma}k_{\nu}   \Big)
\end{equation}
and the Ricci scalar is
\begin{equation}
R = -a^2 (D-1) k^2 \,.
\end{equation}
Finally the Einstein tensor in low energy-momentum mode is
\begin{equation} \label{eq:EinsteinTensorLowpmode}
G_{\mu\nu} = \frac{D-2}{2}a^2 (k^2 \eta_{\mu\nu} - k_{\mu} k_{\nu} ) \,,
\end{equation}
which is proportional to the projection tensor. Therefore, in 4D spacetime, the stress-energy tensor is proportional to the projection tensor,
\begin{equation}
T_{\mu\nu}= \frac{a^2}{8\pi G} (k^2 \eta_{\mu\nu} - k_{\mu} k_{\nu} ) \,. 
\end{equation}

\section{Dynamics of free particle under exponential growth or decay of flat metric}
 Finally we would like to study how a free particle responds when there is exponential growth or decay of the flat background metric. We will study the motion of the test particle using geodesic acceleration similar to the case when one studies gravitational wave \cite{Gwave1, Gwave2}. The explicit formula of geodesic acceleration is given by \cite{Hawking},
\begin{equation} \label{eq:acceleration}
A^{\rho} = \frac{D^2}{D\tau^2} S^{\rho} = R^{\rho}_{\,\,\,\sigma\mu\nu} T^{\sigma}T^{\mu} S^{\nu}\,,
\end{equation}
where $T^{\sigma}$ , $T^\mu$ are tangent vectors to the geodesic, while $S^{\nu}$ is the the normal vector orthogonal to the tangent vectors. Using the Riemannian curvature tensor we have in equation(\ref{eq:CurvatureTensor}), we evaluate the right hand side of (\ref{eq:acceleration}) as
\begin{equation}
\begin{aligned}
&\quad  R^{\rho}_{\,\,\,\sigma \mu\nu} U^{\sigma} U^{\mu} S^{\nu} \\
&= \frac{a^2}{4}\big[ U^{\rho} U^{\sigma}k_{\sigma}k_{\nu} S^{\nu} - S^{\rho}k_{\mu} U^\mu k_{\sigma} U^{\sigma}  \\
& \quad\quad -U_{\sigma}U^{\sigma} k_{\nu} S^{\nu} k^{\rho} + S_{\sigma}U^{\sigma} k_{\mu} U^{\mu} k^{\rho} - k^2 (U^{\rho} U_{\nu} S^{\nu} - S^{\rho} U_{\mu} U^{\mu}  ) \big] \,.
\end{aligned}
\end{equation}
Now we work in the free particle's rest frame that
\begin{equation} \label{eq:4velocityParticleRestframe}
T^{\rho} = U^{\rho} = ( 1,0,0,0 ) \,,
\end{equation}
this would pick out $k_0 $ for the inner product $k_{\mu}U^{\mu}$. And we consider the test particle as a slow moving particle such that we can equate the proper time as the coordinate time $\tau = t$\, then we obtain the acceleration
\begin{equation}
A^{\rho} = \frac{\partial^2 S^\rho}{\partial t^2} = \frac{a^2}{4}\big[ U^{\rho} k_0 k_{\nu}S^{\nu} - S^{\rho} k_0^2 -k_{\nu}S^{\nu} k^{\rho} + S_0 k_0 k^\rho - k^2 (U^\rho S^0 -S^\rho) \big] \,.
\end{equation}
Now first we evaluate the time component $S^0$. Using the fact that $k_0 = k^0 = \omega$, then we have
\begin{equation}
\begin{aligned}
A^0 &= \frac{\partial^2 S^0}{\partial t^2} \\
&=  \frac{a^2}{4} \big[ k_0 k_\nu S^\nu - S^0 \omega^2 - \omega k_{\nu} S^{\nu} + S_0 \omega^2 -k^2 (S^0 -S^0) \big] \\
&= 0 \,.
\end{aligned}
\end{equation}
Hence we have
\begin{equation}
S^0 = vt + c \,,
\end{equation}
where $c$ is a constant of initial position, and $v$ is the velocity along the time direction.
implying that there is no acceleration for the test particle in the time component.
Now we calculate the spatial component, since in the particle rest frame $U^i = 0$,

\begin{equation}
\begin{aligned}
A^i &= \frac{\partial^2 S^i}{\partial t^2} \\
& = \frac{a^2}{4} \big[ - S^i \omega^2 - k_\nu S^{\nu} k^i + S_0 \omega k^i - k^2 ( 0 - S^i)  \big]\\
&= \frac{a^2}{4} \big[ - \omega^2 S^i - (k \cdot S)k^i + c \omega k^i + k^2 S^i  \big] \\
&= \frac{a^2}{4} \big[ (k^2 - \omega^2 ) S^i + S_0\omega k^i - (k \cdot S)k^i \big]
\end{aligned}
\end{equation}
As we have $k^2 - \omega^2 = \omega^2 - |\vec{k}|^2 - \omega^2 = - |\vec{k}|^2  $, then  we obtain the spatial component of the acceleration as,
\begin{equation}
\begin{aligned}
A^i &= \frac{\partial^2 S^i}{\partial t^2} = \frac{a^2}{4} \big[ - |\vec{k}|^2 S^i - (k \cdot S) k^i + c\omega k^i \big] \\
&= \frac{a^2}{4} \big[ - |\vec{k}|^2 S^i - ( k_0 S^0 + k_j S^j  ) k^i + S_0\omega k^i \big] \\
&=  \frac{a^2}{4} \big[ - |\vec{k}|^2 S^i - S_0\omega k^i - k_j S^j   k^i + S_0\omega k^i \big] \\
&=  \frac{a^2}{4} ( - |\vec{k}|^2 S^i - (\vec{k} \cdot \vec{S} )k^i ) \,.
\end{aligned}
\end{equation}
Therefore finally the spatial acceleration is given by
\begin{equation}
A^i = \frac{\partial^2 S^i}{\partial t^2} = -\frac{a^2}{4} (  |\vec{k}|^2 S^i + (\vec{k} \cdot \vec{S} )k^i ) \,.
\end{equation}
The next task is to solve this partial differential equation.  Explicitly we can write down the 3 coupled partial differential equations (PDEs)
\begin{equation}
\begin{cases} 
   \frac{\partial^2 S_x}{\partial t^2} =  -\frac{a^2}{4} \big(  |\vec{k}|^2 S_x + k_x ( k_x S_x + k_y S_y + k_z S_z) \big)  \\
   \frac{\partial^2 S_y}{\partial t^2} = -\frac{a^2}{4}\big(  |\vec{k}|^2 S_y + k_y ( k_x S_x + k_y S_y + k_z S_z) \big) \\
   \frac{\partial^2 S_z}{\partial t^2} = -\frac{a^2}{4} \big(  |\vec{k}|^2 S_z + k_z ( k_x S_x + k_y S_y + k_z S_z) \big)
  \end{cases}
\end{equation}
These three coupled 3 PDEs can be written as in matrix form,
\begin{equation}
 \frac{\partial}{\partial t^2} \begin{pmatrix} S_x \\ S_y \\ S_z \end{pmatrix} = -\frac{a^2}{4} \begin{pmatrix}  
 2 k_x^2 + k_y^2 + k_z^2 & k_x k_y & k_x k_z \\
 k_y k_x & 2 k_y^2 + k_x^2 + k_z^2 & k_y k_z \\
 k_z k_x & k_z k_y & 2k_z^2 + k_x^2 + k_y^2
\end{pmatrix}
\begin{pmatrix} S_x \\ S_y \\ S_z \end{pmatrix}
\end{equation}
where we define the matrix $\mathrm{\pmb{K}}$ as
\begin{equation}
\mathrm{\pmb{K}} = -\frac{a^2}{4} \begin{pmatrix}  
 2 k_x^2 + k_y^2 + k_z^2 & k_x k_y & k_x k_z \\
 k_y k_x & 2 k_y^2 + k_x^2 + k_z^2 & k_y k_z \\
 k_z k_x & k_z k_y & 2k_z^2 + k_x^2 + k_y^2
\end{pmatrix}
\end{equation}
To solve this matrix equation, first we assume the solution takes the following form,
\begin{equation}
S_i ( \vec{x} ,t ) = A_i v_i e^{(\Omega t - \vec{\kappa} \cdot \vec{x}  ) } \,,
\end{equation}
where $A_i$ are constants and $v_i$ are eigenvectors. It is noted that the angular frequency $\Omega$ here is different from that of the original $\omega$ in the phase factor, the $\Omega$ is the angular frequency of the test particle induced from the flat spacetime oscillation, and the same idea goes for the difference between wave vector $\pmb{\kappa}$ and $\mathrm{\pmb{k}}$. Then substituting the solution to the matrix from above, we obtain
\begin{equation}
\begin{pmatrix} \Omega^2 B_x \\ \Omega^2 B_y \\\Omega^2 B_z \end{pmatrix} = -\frac{a^2}{4} 
\begin{pmatrix}  
 2 k_x^2 + k_y^2 + k_z^2 & k_x k_y & k_x k_z \\
 k_y k_x & 2 k_y^2 + k_x^2 + k_z^2 & k_y k_z \\
 k_z k_x & k_z k_y & 2k_z^2 + k_x^2 + k_y^2
\end{pmatrix}
\begin{pmatrix} B_x \\ B_y \\ B_z \end{pmatrix} \,.
\end{equation}
Then we can solve it by requiring
\begin{equation}
{\mathrm{det}} 
\begin{pmatrix}  
 \Omega^2 +  \frac{a^2}{4} (2 k_x^2 + k_y^2 + k_z^2 ) & \frac{a^2}{4} k_x k_y & +\frac{a^2}{4} k_x k_z \\
 \frac{a^2}{4} k_x k_y &  \Omega^2 +  \frac{a^2}{4}  (2 k_y^2 + k_x^2 + k_z^2 ) & \frac{a^2}{4} k_y k_z \\
 \frac{a^2}{4} k_x k_z & \frac{a^2}{4} k_y k_z &  \Omega^2 +  \frac{a^2}{4} (2 k_z^2 + k_x^2 + k_y^2 )
\end{pmatrix}
=0
\end{equation}
which is to solve
\begin{equation}
\mathrm{det} ( \Omega^2 \mathrm{\pmb{I}} + \mathrm{\pmb{K}}  ) =0
\end{equation}
and as if for the eigenvalues as
\begin{equation}
\mathrm{det} ( - \lambda \mathrm{\pmb{I}} + \mathrm{\pmb{K}} ) =0 \,.
\end{equation}
Then we can relate the two by
\begin{equation}
\Omega = \pm i \sqrt{\lambda} \,.
\end{equation}
Upon solving there are three eigenvalues of the $\mathrm{\pmb{K}}$ matrix, for which all depend on the flat spacetime oscillation plane wave frequency and wave vector,
 we have
 \begin{equation}
 \lambda_1 = \lambda_2 =\frac{a^2}{4}( k_x^2 + k_y^2 + k_z^2 ) \, \, , \,\,  \lambda_3 = \frac{a^2}{2}( k_x^2 + k_y^2 + k_z^2 ) 
 \end{equation}
Therefore
\begin{equation}
\Omega_1 = \Omega_2 = \pm \frac{ia}{2} | \vec{k} | \,\,, \,\, \Omega_3 =\pm \frac{ia}{\sqrt{2}} | \vec{k} | \,.
\end{equation}
And the corresponding three orthogonal eigen-vectors are
\begin{equation}
v_1 = ( -k_z , 0 , k_x ) \,\,,\,\, v_2 = (-k_y , k_x ,0 ) \,\,,\,\, v_3 = ( k_x , k_y , k_z) \,.
\end{equation}
Thus, we obtain the exact solution for the coupled PDEs. We can see that there is no acceleration in the $S^0$ time component, while for the the $S^i$ spatial components, it is oscillatory.

\section{Conclusion}
In this paper, we have analysed the geometry of the conformal metric. The geodesic equation is studied. The stress-energy tensor is studied with the perfect fluid model, obtaining positive matter density and negative spacetime pressure for normal matter; negative matter density and positive pressure for virtual light-like matter; and both zero density and pressure for massless matter. Then we study the geometry of the conformal metric in low energy-momentum regime, for which the stress-energy tensor is proportional to the projection tensor. Finally, we solve for the exact solution of the geodesic deviation equation for the conformal metric.

\end{document}